\documentclass[aps,prl,nofootinbib,twocolumn,showpacs,superscriptaddress,letterpaper,amsmath,amssymb]{revtex4}

\usepackage{graphicx}  
\usepackage{dcolumn}   
\usepackage{bm}        
\usepackage{amssymb}   
\usepackage{feynmf}
\usepackage{slashed}
\def\gsim{\lower0.5ex\hbox{$\:\buildrel >\over\sim\:$}}
\def\lsim{\lower0.5ex\hbox{$\:\buildrel <\over\sim\:$}}

\newcommand{\be}{\begin{equation}}
\newcommand{\ee}{\end{equation}}
\newcommand{\bea}{\begin{eqnarray}}
\newcommand{\eea}{\end{eqnarray}}

\newcommand{\nbox}{{\,\lower0.9pt\vbox{\hrule \hbox{\vrule height 0.2 cm
\hskip 0.2 cm \vrule height 0.2 cm}\hrule}\,}}

\begin{document}

\thispagestyle{empty}
\vspace*{-3.5cm}

\vspace{0.5in}

\title{Two Lines or Not Two Lines? That is the Question of Gamma Ray Spectra}

\begin{center}
\begin{abstract}
Lines in the spectrum of cosmic gamma rays are considered one of the more robust 
signatures of dark matter annihilation.  We consider such processes from
an effective field theory vantage, and find that generically, two or more lines
are expected, providing an interesting feature that can be exploited for searches
and reveal details about the underlying theory of dark matter.  Using the 130 GeV
feature recently reported in the Fermi-LAT data as an example, we analyze the
energy spectrum in the multi-line context and find the data to be consistent
with a single $\gamma\gamma$ line, a single $\gamma Z$ line or both a $\gamma \gamma$ and a $\gamma Z$ line.
\end{abstract}
\end{center}

\author{Arvind Rajaraman}
\author{Tim M.P. Tait}
\author{Daniel Whiteson}
\affiliation{Department of Physics and Astronomy, University of California, Irvine, CA 92697}
\preprint{UCI-HEP-TR-2012-XX}
\maketitle

\section{Introduction}

Nonbaryonic dark matter is now a crucial element in the picture of the cosmology 
of the early Universe.  And yet, its role
in the framework of particle physics has remained elusive.  Discovery
of any kind of non-gravitational
interactions of dark matter, through observation of its production at high energy 
accelerators, its direct scattering
with heavy nuclei, or its annihilation is an area of major
experimental activity.  Any of these observations would establish dark 
matter as an exotic particle, and would give insights into its nature.

In the search for WIMP annihilation, gamma rays provide a promising window.  
Gamma rays produced in the galaxy do not typically scatter on their way to the Earth, 
providing a handle from the morphology of their origin.
If dark matter annihilates
into quarks (or any particle with large decay branching ratios into quarks, such as 
$W$, $Z$, and Higgs bosons), the resulting
spectrum of gamma rays tends to be rather soft, arising from the eventual decays 
of $\pi^0$'s produced in the hadronic showers, and with a cut-off at the mass of the 
WIMP.  These continuum signals are difficult to extract 
from the (often unknown)
astrophysical backgrounds, and so to date
searches have been most efficient when observing 
regions of the sky
which are largely background free, such as dwarf spheroidal 
galaxies \cite{Ackermann:2011wa}.

If dark matter annihilates into charged particles, it must also be able to annihilate 
directly into two-body final states including a photon.  Such processes are mediated by 
loops, and thus are suppressed compared to the
continuum annihilation.  Their power comes from the feature that they 
produce a photon with a well-defined
energy given (for the process $\chi \chi \rightarrow \gamma Y$) by
\bea
E_\gamma = m_\chi \left( 1 - \frac{M_Y^2}{4 m_\chi^2} \right)
\label{eq:eline}
\eea
where $M_Y$ is the mass of the second annihilation product.  For the case 
where $Y=\gamma$, the line occurs
at an energy equal to the mass of the WIMP itself, $E_\gamma=m_\chi$.  Given this striking feature, the 
search for gamma ray lines has become a standard item on the menu of searches for 
WIMP annihilation using Fermi-LAT data \cite{Abdo:2010nc,Fermi:2012}.

While it is possible for instrumental effects or more prosaic astrophysical 
processes \cite{Profumo:2012tr}
to mimic a bump in the gamma ray spectrum, a line remains one of the most 
compelling prospects for
the indirect detection of dark matter annihilation.  Motivated by the recent tentative 
indication that there may be
such a feature at an energy around 130 GeV \cite{Bringmann:2012vr,Weniger:2012tx}
with a relatively large (rough $1/10$ of the thermal expectation) cross section, and 
consistent with originating close to the
galactic center \cite{Bringmann:2012vr,Weniger:2012tx,Tempel:2012ey}
we explore the generic properties that one might expect in a theory which can 
produce strong line signals.  We use the Fermi-LAT data
as analyzed in Ref.~\cite{Weniger:2012tx} to illustrate how one may dissect a 
putative line signal both to lend strength
to its origin from dark matter annihilations, as well as to learn something 
about the details of the theory of dark
matter.

\section{The Theory Space of $\gamma$ Ray Lines}

Boiled down to its essence, the process $\chi \chi \rightarrow \gamma Y$ results from an amplitude
involving a loop of charged
particles which also couple to $Y$.  The charged particles in the loop could be either exotic 
states, or
part of the Standard Model, or (as is typical) a mixture of the two.
For the current discussion,
rather than wed ourselves to any specific UV-complete theory, we work in an effective theory
framework and discuss operators in the effective action allowing WIMPs to annihilate into two particle final states, one of which is a $\gamma$-ray.

The operator description is only appropriate to
describe theories for which the momentum transfer is smaller 
than the masses of any of
the mediators which have been integrated out.  For annihilation, the momentum transfer
is $\sim m_\chi$, so this restriction boils down to the requirement
that all of the charged loop particles are much heavier than the WIMP itself (but one
can enlarge the effective theory to capture cases where some of the loop particles
are heavier than the WIMP, and some are SM particles, see 
Ref.~\cite{Goodman:2010qn} for an example and Ref.~\cite{Abazajian:2011tk}
for some related discussion).

In constructing the operators in the EFT language, we consider both scalar
and fermionic WIMPs.  We identify the leading operators of each type, and ignore
higher mass dimension operators which are presumably further suppressed
by the heavy mediator masses.  We work in a description where the
$SU(2)_L \times U(1)_Y$ gauge symmetry is realized manifestly.  In counting the
dimension of operators, this choice is equivalent to the assumption that the
charged particles which have been integrated out have masses largely
independent of electroweak symmetry-breaking.

The natural ingredients from which operators are built are the field strengths
of the hypercharge and $SU(2)$ gauge fields, $B_{\mu \nu}$
and $W^a_{\mu \nu}$, the Higgs doublet $\Phi$,
and the dark matter field $\chi$ (which we take to be either
a scalar or spin-$1/2$ fermion). We will  build operators
up to dimension-7 out of these ingredients, focusing on
operators which produce at least one photon.  We classify each operator
according to what type(s) of line process(es) it mediates, including
$\gamma \gamma$, $\gamma Z$ 
\cite{Bergstrom:1997fh,Gustafsson:2007pc,Bergstrom:2004nr,cline:2012},
and/or $\gamma h$  \cite{Jackson:2009kg,Bertone:2010fn}
(where $h$ is the SM Higgs boson)\footnote{We necessarily will miss annihilation
into a photon together with a dark sector particle \cite{Bertone:2009cb},
which would require extension of the effective theory to contain the
second decay product.}.  We further distinguish operators leading to
velocity-suppressed or -unsuppressed rates into gamma ray lines
depending on whether the leading term in the expansion of 
$\langle \sigma v \rangle$ for small relative WIMP velocity $v$ is
a constant or is proportional to $v^2$.  Since $v \sim 10^{-3}$ in the
Milky way halo, velocity-suppressed operators need much larger
couplings in order to produce a visible signal to compensate their
$v^2$ suppression.

At dimension-4, for either a scalar or fermionic WIMP, the unique choice
leading to coupling to the photon requires that the dark matter particles themselves
are charged under $U(1)_Y$ and/or $SU(2)_L$ leading after EWSB
to couplings to $\gamma$ and $Z$.   The cosmological
bounds on such ``milli-charged" WIMPs
are very strong \cite{McDermott:2010pa}, leaving this possibility unlikely to
produce a line feature that could be observable by any near future experiment.

At dimension-5, there are two operators for a Dirac fermion
\bea
\bar{\chi} \gamma^{\mu\nu} \chi ~ B_{\mu \nu} ~~{\rm and}~~
\bar{\chi} \gamma^{\mu\nu} \chi  ~ \tilde{B}_{\mu \nu}
\eea
where $\gamma^{\mu \nu} \equiv 1/4 [ \gamma^\mu,\gamma^\nu]$.  These 
operators correspond to a weak magnetic (electric)
dipole moment for the WIMP and leads
to unsuppressed annihilation into both $\gamma \gamma$ and $\gamma Z$.

At dimension-6, there is a family of operators built out of the set of dimension-4
factors,
\bea
\left\{
B_{\mu\nu}B^{\mu\nu},~~
W^{a}_{\mu\nu}W^{a\mu\nu},~~
B_{\mu\nu}\tilde{B}^{\mu\nu},~~
W^{a}_{\mu\nu}\tilde{W}^{a\mu\nu}
\right\}
\label{eq:F2}
\eea
multiplied by $\chi^2$ (or $|\chi|^2$ if $\chi$ is a complex scalar).  These operators
also lead to unsuppressed annihilation into 
both $\gamma \gamma$ and $\gamma Z$.

Similarly, at dimension-7, there is another  family of operators built out of the same
set~(\ref{eq:F2}) of dimension-4
factors, multiplied by
\bea
\bar{\chi}\chi ~~~{\rm or} ~~~ \bar{\chi}\gamma_5\chi ~.
\eea
for either a Majorana or Dirac fermion.  These operators generate
both $\gamma \gamma$ and $\gamma Z$ annihilations which
are unsuppressed (suppressed) for $\bar{\chi} \gamma_5 \chi$
($\bar{\chi} \chi$).

Finally, for a Dirac fermion, we can also have the dimension-7 terms
involving  tensor operators formed from $\bar{\chi}\gamma^{\mu\nu}\chi$
multiplied by a factor from either the set,
\bea
\left\{
B_{\mu\alpha}\tilde{B}^{\alpha\nu},~~
W^a_{\mu\alpha}\tilde{W}^{a\alpha\nu}\right\},
\label{eq:tensorF2}
\eea
or
\bea
\left\{
B_{\mu\nu}|\Phi|^2, ~~\tilde{B}_{\mu\nu}|\Phi|^2,~~
\Phi^\dagger {W}^a_{\mu\nu}T^a\Phi,~~ \Phi^\dagger \tilde{W}^a_{\mu\nu}T^a\Phi
\right\},
\label{eq:tensorF1}
\eea
the set contained in (\ref{eq:tensorF2}) leads again to both
$\gamma \gamma$ and $\gamma Z$ lines
(unsuppressed), whereas the set contained in 
(\ref{eq:tensorF1}) leads to a single unsuppressed $\gamma h$ line.

At dimension-8, there is a very large number of operators, which we
will not catalogue exhaustively. 
We note, however, that at dimension-8, there are operators 
built out of the vector currents,
$J_\mu 
= (\chi\partial_\mu \chi^*-\chi^*\partial_\mu \chi)$~(for a complex scalar WIMP),
$S_\mu = \chi\gamma_\mu\chi$
for a Dirac fermion WIMP, and
$S^5_\mu = \chi\gamma_\mu\gamma_5\chi$ 
for either a Dirac or Majorana fermion WIMP.  Any of them may be combined with
a factor from the set,
\bea
& &
\left\{ 
B_{\mu\alpha}\Phi^\dagger D_\alpha \Phi,~~
\tilde{B}_{\mu\alpha}\Phi^\dagger D_\alpha \Phi,~~
\Phi^\dagger W^a_{\mu\alpha}T^a D_\alpha \Phi,
\right.
\nonumber \\
& &
\left. ~~~~~~~~
\Phi^\dagger \tilde{W}^a_{\mu\alpha}T^a D_\alpha \Phi
\right\}.
\eea
All operators in this set lead to both $\gamma Z$ and $\gamma h$ lines.
Those constructed from $J_\mu$ and $S^5_\mu$ are $v$-suppressed,
whereas $S_\mu$ is unsuppressed.
These operators are naturally generated in models where the dark matter
dominant communication with the SM particles is through exchange of a
$Z^\prime$ boson (for examples of models where this
is the case, 
see \cite{Dudas:2009uq,Mambrini:2009ad,Jackson:2009kg,Dudas:2012pb}).  
The absence of a $\gamma \gamma$ line is naturally explained
by the Landau-Yang theorem \cite{Yang:1950rg}, which forbids a neutral vector 
state from decaying into two photons.

The simple exercise of cataloguing operators in the effective field theory
already reveals very interesting features.  Every operator considered leads to
two lines ($\gamma \gamma$ and $\gamma Z$) or $\gamma Z$ and
$\gamma h$), with a simple prediction for their
relative intensities.
The sole exception is the set of operators contained in (\ref{eq:tensorF1})
which exist only for a Dirac WIMP and lead to a single $\gamma h$ line.

In terms of the underlying picture in which line processes are mediated
by charged particles running in a loop, the fact that there are multiple
lines corresponds to the fact that such particles must be charged
under $SU(2)_L$ and/or $U(1)_Y$ and thus must couple to both the
photon and the $Z$ boson.  Similarly, coupling to a Higgs boson
often is accompanied by coupling to a longitudinal $Z$ boson (but not always --
if the coupling is to $\Phi^\dagger \Phi$, it implies interactions with
one or two Higgs bosons, but always two longitudinal $Z$ bosons).
Presumably any realistic UV complete theory will generate more than one
operator, allowing for the possibility of interference.
Interference will adjust the relative sizes of the two or three lines, but in the
absence of fine tuning will not cancel one of them completely.

Our results suggest interesting variations in the experimental analyses
of gamma ray lines.  First, one may search
for two (or three) lines whose energies are consistent with
Eq.~(\ref{eq:eline}) for a single WIMP mass.  Such an observation
would be highly suggestive of dark matter annihilation, and less likely to 
be produced by astrophysical or instrumental effects.  Of course, the ability 
to resolve two lines is very challenging for WIMP
masses larger than $\sim 150$ GeV, because of the finite energy resolution 
of the detector ($\Delta E / E \sim 10\%$ at $E \sim 100$~GeV for the Fermi-LAT).
Second, if there is a concrete observation of a single line, will be highly suggestive
of a Dirac fermion WIMP annihilating through one of a definite set of
effective interactions, providing clues to the nature of the UV theory.

\section{The Fermi 130 GeV Feature as a Case Study}

In order to explore how multiple lines could manifest themselves in realistic data,
we analyze the observed $\gamma$-ray spectrum of  Fermi-LAT, including the feature at
$E_\gamma=130$ GeV~\cite{Weniger:2012tx}, in
the context of the multi-line theory.  While it is premature to interpret this feature
as dark matter annihilation, it provides an interesting case study which could
even turn out to ultimately tell us something about dark matter.
We use the regions of interest
found to have largest significance, Reg3 and 
Reg4 \cite{Bringmann:2012vr,Weniger:2012tx}
and the ULTRACLEAN
photon selection.  While not presented here, we have also performed the
analysis for the looser photon selection (SOURCE class), which yields very
similar results.
We will focus our analysis on the $\gamma \gamma$ plus
$\gamma Z$ case; the extrapolation for $\gamma Z$ plus
$\gamma h$ is straightforward.

We follow the standard Fermi analysis procedure, evaluating the relative likelihood of the background-only hypothesis (null) and the background-plus-signal 
(best) hypothesis using the test statistic ($TS$):
\bea
TS &=& -2 \rm{ln} \frac{ \mathcal{L}_{\rm null} }{ \mathcal{L}_{\rm best}}
 \eea
but with a likelihood $\mathcal{L}$ which includes both a 
power-law background
model as well as terms for potential $\gamma\gamma$ and $\gamma Z$ lines:
\bea
\mathcal{L}(E_\gamma | N_{\gamma\gamma}, N_{\gamma Z}, \beta,\alpha) =
~~~~~~~~~~~~~~~~~~~~~~~~~~~~~
\nonumber
\eea
\vspace*{-0.75cm}
\bea
& &
  \beta\left( \frac{E_\gamma}{E_0}\right) ^{-\alpha} 
~ +  ~~ N_{\gamma\gamma}~f_{\rm DM}(E_\gamma | m_\chi) \nonumber \\
& & + ~~~~ N_{\gamma Z}~f_{\rm DM} \left(E_\gamma | m_\chi 
\left( 1 - \frac{M_Z^2}{4 m_\chi^2} \right) \right)
\eea
The function $f_{DM}(E_\gamma | E_{\rm  line})$ is a normalized double Gaussian function fit to the expected line
shape for $E_{\rm line}=130$ GeV as provided in
Ref~\cite{Weniger:2012tx}. For other values of the expected peak location $E_{\rm line}$, the values of
the Gaussian widths and means are treated as linearly dependent on the
position of the expected peak.  The parameters $N_{\gamma\gamma}$ and
$N_{\gamma Z}$ control the total yield from the $\gamma\gamma$ and $\gamma
Z$ processes, respectively.  The two terms
describe the correlated $\gamma\gamma$ and $\gamma Z$ contributions. 

For $\mathcal{L}_{\rm best}$, the parameters  
($N_{\gamma\gamma}, N_{\gamma Z},\beta,\alpha$)  are floated to find the 
maximum likelihood value. For $\mathcal{L}_{\rm null}$, the yields 
($N_{\gamma\gamma}, N_{\gamma Z}$) are fixed to zero and the
background model parameters ($\beta,\alpha$) are floated to their
best fit values. The local statistical significance may be interpreted in the asymptotic
regime~\cite{fermistat}, as $\sigma = \sqrt{TS}$.

\begin{figure}
\includegraphics[width=3in]{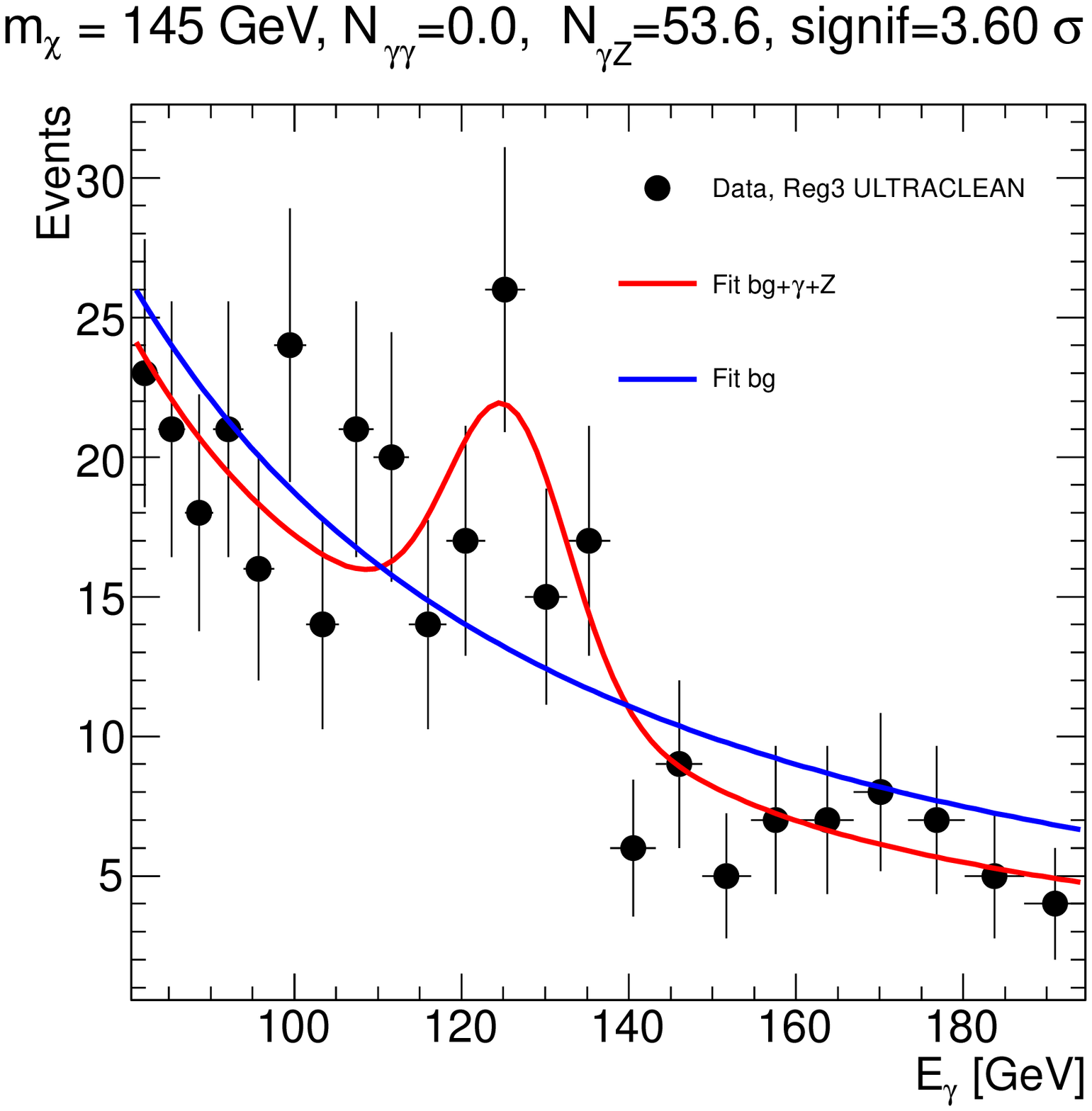}\\
\includegraphics[width=3in]{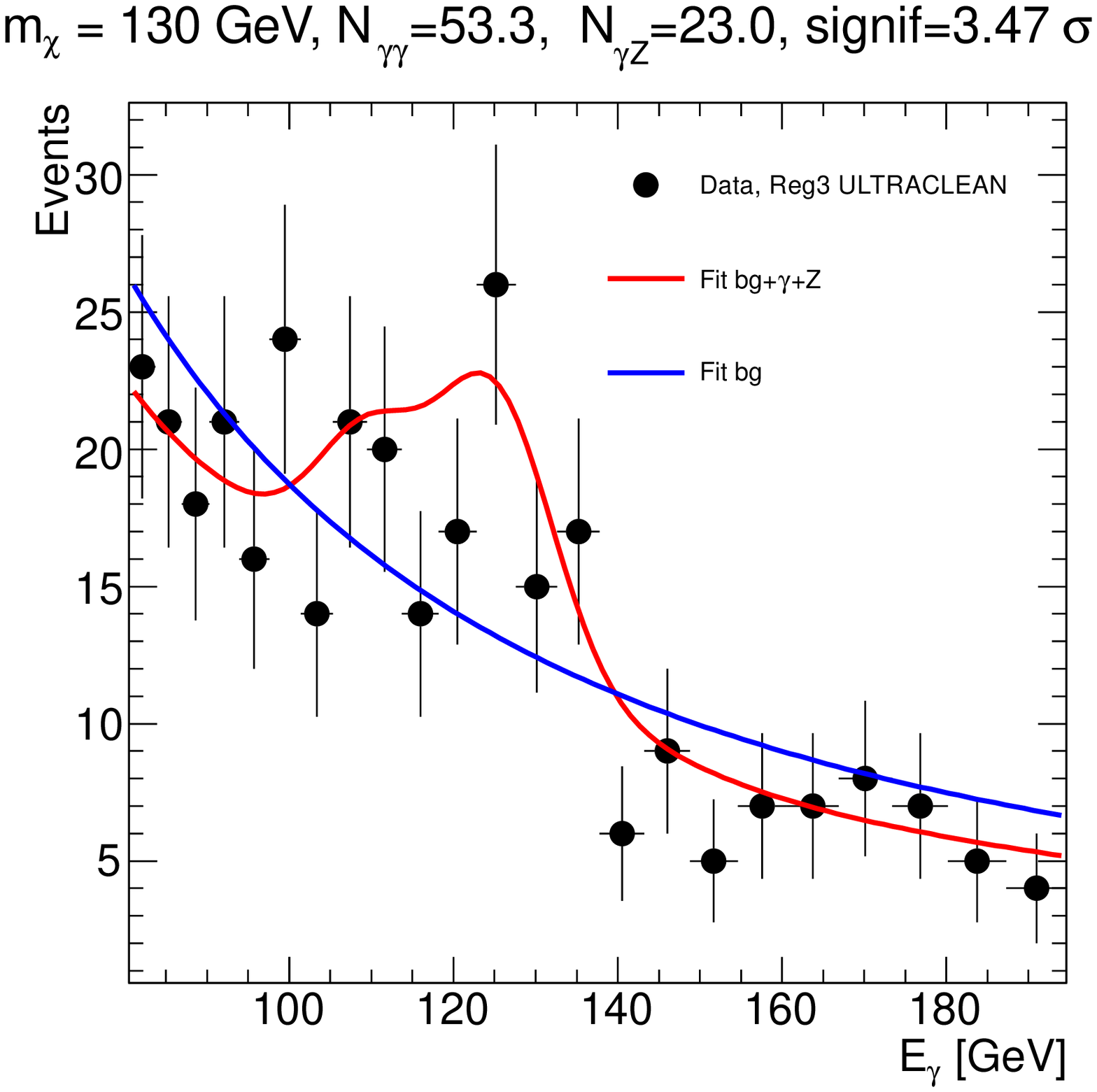}
\caption{Sample spectrum predicted by dark matter annihilation into a single $\gamma Z$ 
line (top panel)
or two lines produced by $\gamma \gamma$ and $\gamma Z$ (bottom panel).  
Data from Ref.~\cite{Weniger:2012tx}
is overlaid.
\label{fig:spectra}}
\end{figure}

Example fits are shown in Fig~\ref{fig:spectra}. If the WIMP is
assumed to have mass of 145 GeV, a $\gamma Z$ process would produce a
line at $E_\gamma=130$ GeV. In this case, the feature at
$E_\gamma=130$ GeV can be interpretted as pure $\gamma Z$; any
contribution from $\gamma\gamma$ would appear at larger $E_\gamma$,
where no such  feature appears.   If, on the other hand, the WIMP is
assumed to have a mass of 130 GeV, then a $\gamma\gamma$ process would
produce a line at $E_\gamma=130$ GeV, explaining the 
feature\footnote{In Ref.~\cite{Tempel:2012ey} it is claimed that in order to produce
a peak at $E_\gamma=130$ GeV after electromagnetic showering
of the parent photon, a WIMP mass of $\sim 145$~GeV is required.
We disagree with this statement.}. In this
case, however, there is room at lower values of $E_\gamma$ for
contributions from a $\gamma Z$ process. The two fits have
approximately equal signifiance.

\begin{figure}
\includegraphics[width=3in]{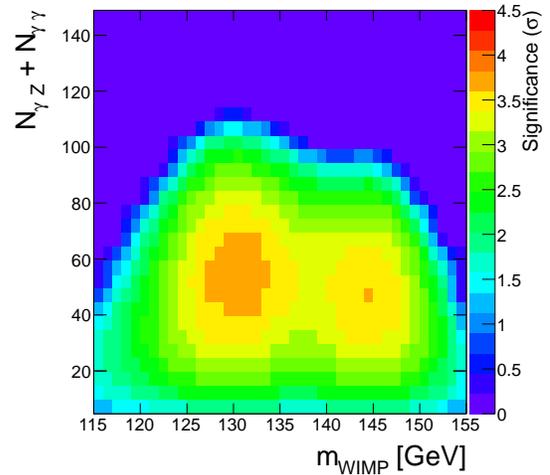}
\includegraphics[width=3in]{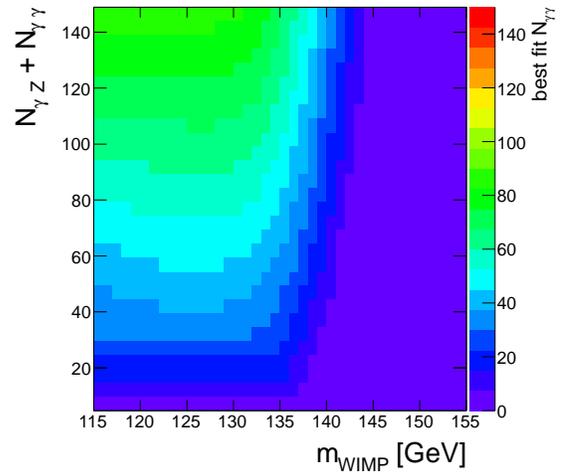}
\caption{Top, local statistical significance of the signal versus WIMP mass and total yield
  due to $\gamma\gamma$ and $\gamma Z$ lines in the Reg3 region of
  interest defined in Ref.~\cite{Weniger:2012tx}. Bottom, the contribution
  from $\gamma\gamma$ at each point in the plane. 
  \label{fig:rgn3sum}}
\end{figure}

\begin{figure}
\includegraphics[width=3in]{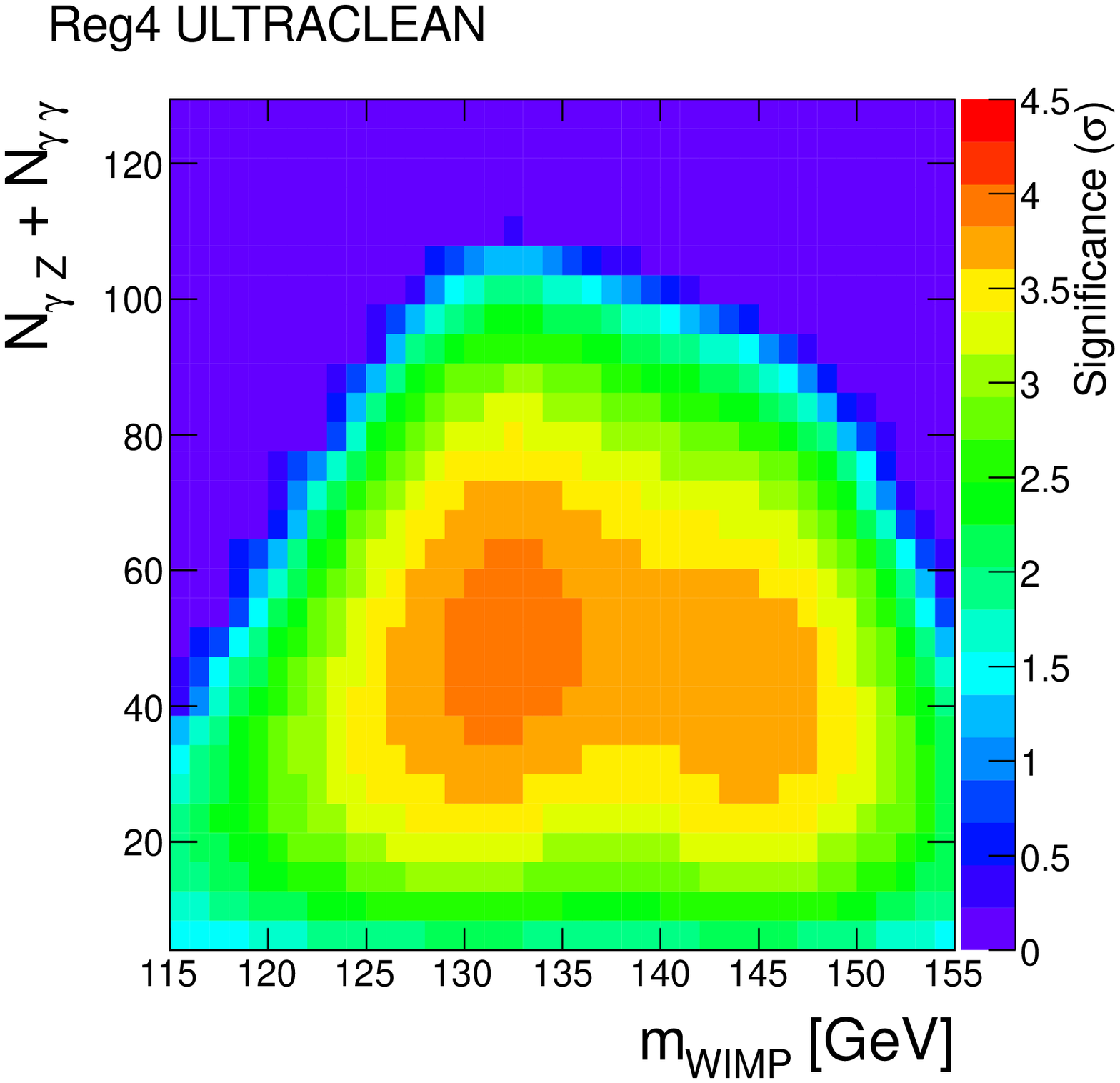}
\includegraphics[width=3in]{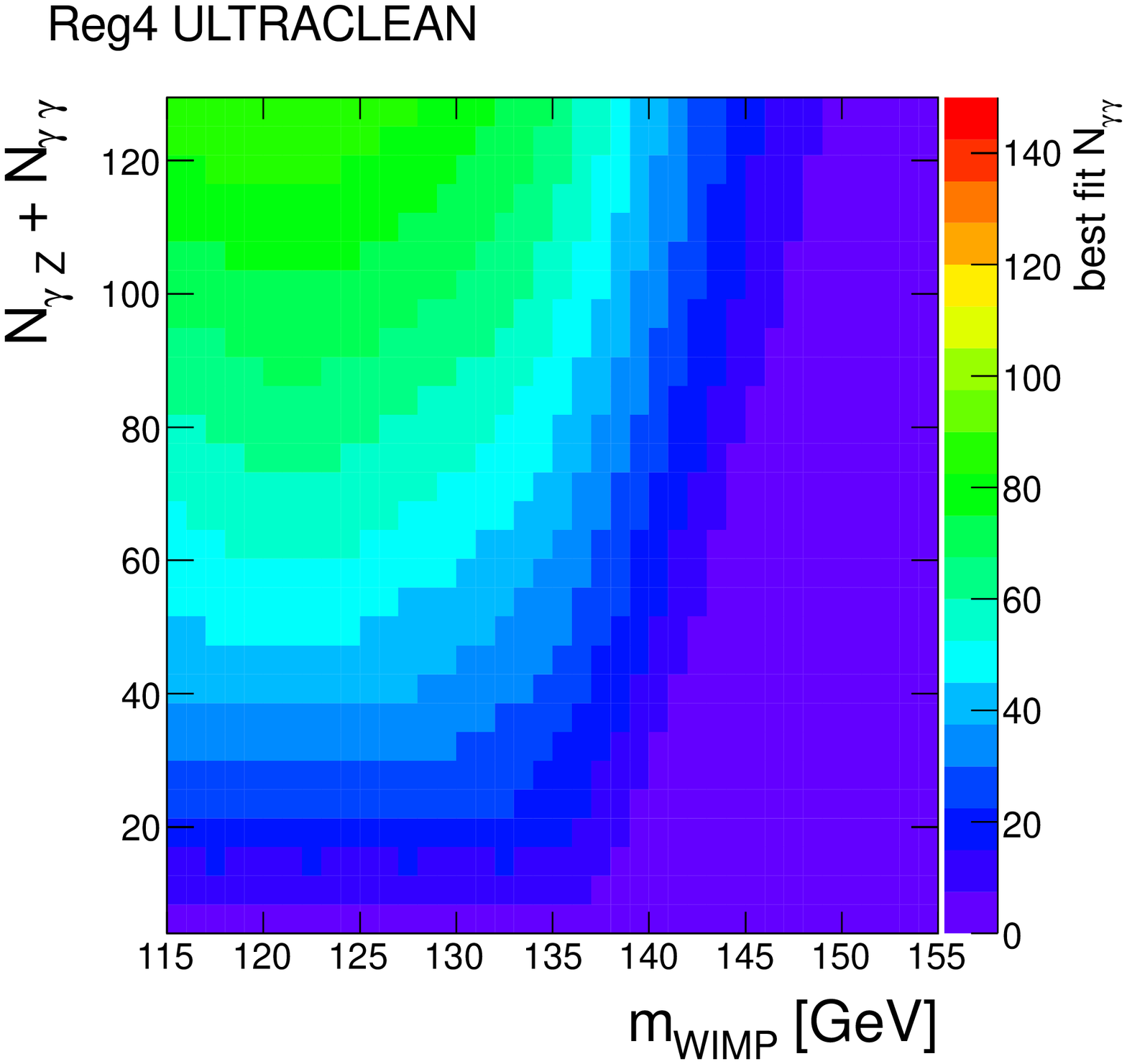}
\caption{Top, local statistical significance of the signal versus WIMP mass and total yield
  due to $\gamma\gamma$ and $\gamma Z$ lines in the Reg4 region of
  interest defined in Ref.~\cite{Weniger:2012tx}. Bottom, the contribution
  from $\gamma\gamma$ at each point in the plane.
\label{fig:rgn4sum}}
\end{figure}

Figure~\ref{fig:rgn3sum} shows a scan of WIMP masses and total yields
($N_{\gamma\gamma}+N_{\gamma Z}$), revealing the two regions of
maximal significance, near $m_{\chi}=130$ and 145 GeV. At each
point, the contribution from $\gamma\gamma$ is also shown.  If $m_{\rm
  WIMP} = 130$ GeV, then the interpretation is consistent with a large
$\gamma\gamma$ contribution, but cannot rule out some contribution
from $\gamma Z$.  If, however, $m_{\chi}=145$ GeV, then a pure
$\gamma Z$ interpretation is preferred. Figure~\ref{fig:rgn4sum} shows the same 
scan for Reg4, which has largely the same features.

\begin{figure}
\includegraphics[width=3in]{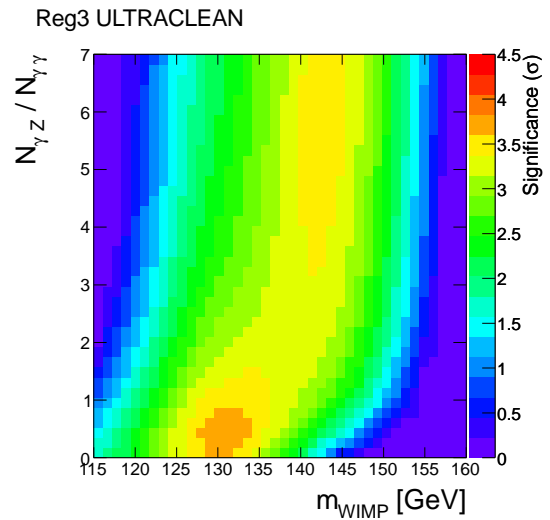}
\includegraphics[width=3in]{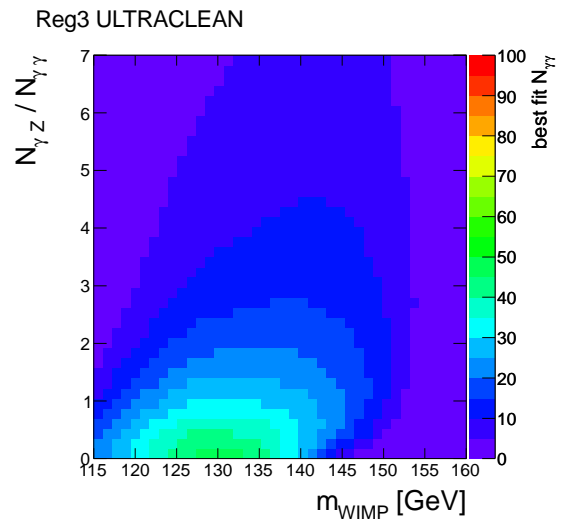}
\caption{Top, local statistical significance of the signal versus WIMP
  mass and ratio of yields in the $\gamma\gamma$ and $\gamma Z$ lines in the Reg3 region of
  interest defined in Ref.~\cite{Weniger:2012tx}. Bottom, the contribution
  from $\gamma\gamma$ at each point in the plane.
\label{fig:rgn3ratio}}
\end{figure}

\begin{figure}
\includegraphics[width=3in]{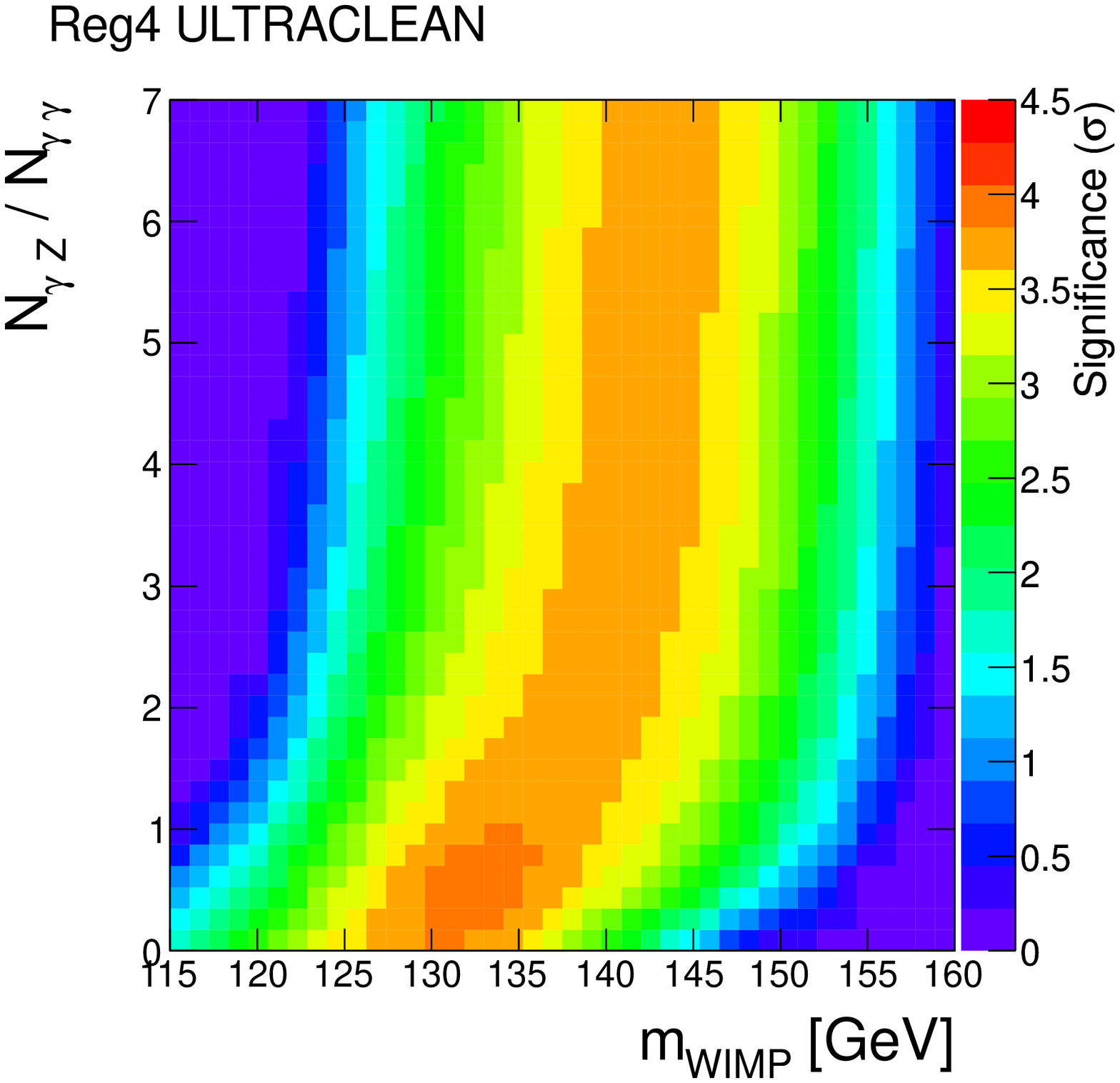}
\includegraphics[width=3in]{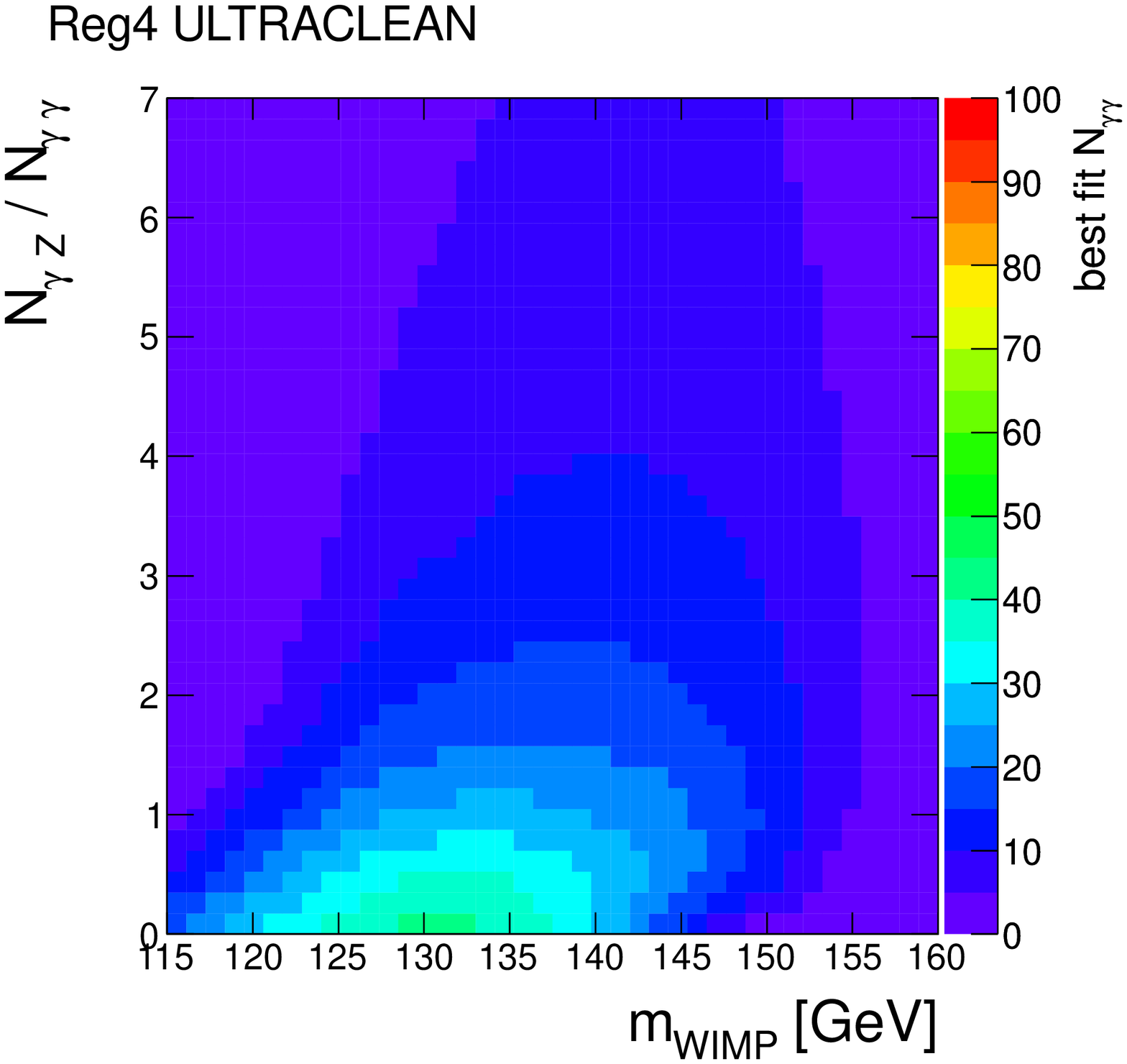}
\caption{Top, local statistical significance of the signal versus WIMP
  mass and ratio of yields in the $\gamma\gamma$ and $\gamma Z$ lines in the Reg4 region of
  interest defined in Ref.~\cite{Weniger:2012tx}. Bottom, the contribution
  from $\gamma\gamma$ at each point in the plane.
\label{fig:rgn4ratio}}
\end{figure}

In Figures~\ref{fig:rgn3ratio} and \ref{fig:rgn4ratio} the scans are performed
as a function of WIMP mass and the ratio of $\gamma Z$ to
$\gamma\gamma$ yields for Reg3 and Reg4.  Note that while the region of
maximum significance is near $m_{\chi}=130$ GeV, it prefers
$N_{\gamma Z}/N_{\gamma\gamma}>0$.  Also shown in each figure is the
value of $N_{\gamma \gamma}$ corresponding to the best fit for each
point in ($m_\chi,N_{\gamma Z}/N_{\gamma\gamma}$).

The two regions show consistent features.  The maximum significance is consistent 
with either a pure $\gamma Z$ or pure  $\gamma\gamma$ scenario; the 
interpretation of the $E_\gamma=130$ GeV line in the $\gamma\gamma$ scenario 
also allows for $\gamma Z$ contributions at lower $E_\gamma$.  In fact, the best
fits have a non-zero fraction of $\gamma Z$ (less than one), 
but this preference is not very significant.  Clearly more data would be very helpful
in terms of sharpening this analysis in order to draw more firm conclusions from it.

From Ref.~\cite{Weniger:2012tx},
the best fit value of the cross section (assuming a $\gamma \gamma$ interpretation)
is about $10^{-27}$~cm$^3$/s $\sim 10^{-4}$~TeV$^{-2}$
for dark matter distributed according to an
NFW profile.  From here one could
compare with detailed calculations based on the operators in the effective
field theory catalogue to determine a consistent parameter space, but we leave
such detailed comparisons for future work, and instead interpret such a target
cross section schematically.

Focusing as an example on any one of the dimension-6 operators for scalar 
WIMPs, we normalize the operator as
$\alpha \alpha_\chi / M^2$, where $\alpha \equiv e^2 / (4 \pi)$ is the
electromagnetic coupling and $\alpha_\chi \equiv g^2 / (4 \pi)$ represent
(unknown) couplings in the dark sector.  This choice of normalization is consistent
with the operator being generated at one loop, with $M$ playing the role of the
mass(es) of the particles in the loop.  Obviously, this implementation is subject to
unknown numerical factors such as the number of species contributing inside
the loop, as well as factors associated with their spins, etc.  The idea is to get
a very rough sense for the mass scale $M$ of the loop particles, given the
target cross section of the Fermi feature.

Our simple estimate indicates that provided there is no velocity-suppression,
\bea
M & \sim & \sqrt{\alpha_\chi}~ 150~{\rm GeV}.
\eea
This is an interesting result.  If the dark sector is strongly coupled 
($\alpha_\chi \sim 1$), the loop particles should have masses in the
range of 150 GeV, safely above the LEP bound 
of about 100 GeV, but low enough that
the LHC has an opportunity to observe them through electroweak production.
For weaker $\alpha_\chi$, the mass must be lower to compensate, rapidly coming
into conflict with the LEP bound for $\alpha_\chi \lesssim 0.5$.  

For a velocity-suppressed operator, the target mass is of the order 
$M \sim \sqrt{\alpha_\chi}~ 5~{\rm GeV}$, far enough below the LEP
bound that not even a strongly coupled dark sector would be able to reconcile
the two.  However, it is worth mentioning a few provisos to this statement.
For example,
one way in which the EFT could spectacularly break down would be when 
there is an additional dark sector state which appears in the $s$-channel for
annihilation.  Very large enhancements are possible in this case, depending how
close to on-shell the resonance is for WIMP annihilation.
In addition, if a large multiplicity of species contribute to the line annihilation,
the amplitude will grow with the number.  For a rather
extreme multiplicity of $\sim 500$,
a $v^2$-suppressed annihilation would be consistent with the LEP bound
for $\alpha_\chi \sim 1$.

\section{Outlook}

Annihilation of dark matter into a two body final state containing a photon provides
a striking signature, and is one of the most promising prospects for an indirect
detection of dark matter.  In this article, we have explored some generic
features of gamma ray lines using an effective theory framework.

The effective theory illustrates a fascinating feature -- the operators which give
rise to one gamma ray line, typically also give rise to two.  For a scalar or
Majorana WIMP, every operator considered produces either $\gamma \gamma$
{\em and} $\gamma Z$, or $\gamma Z$ {\em and} $\gamma h$, and
the intensities of each line are correlated for a given operator.  Multiple lines
are a {\em generic} feature, and one that can be used to improve searches
in data from gamma ray observatories, or help match to specific UV complete
theories once a discovery is made.  For a Dirac WIMP, one class of operators
provides an exception to the multiple-line rule, producing a single $\gamma h$
line.  Nonetheless, observation of
a single line provides very specific information about the
nature of the theory of dark matter.

Using the recent observation of a feature at 130 GeV in the Fermi-LAT data, we
analyze the data in a multi-line context, and find that there is a very
mild preference for contribution from
two lines, though uncertainties are large.
Should this feature persist and not 
ultimately prove to be instrumental or astrophysical in nature, more data should
help sharpen this analysis and make more concrete statements.

\acknowledgements
\section{Acknowledgements}

DW is supported by grants from the Department of Energy
Office of Science and by the Alfred P. Sloan Foundation. The work of 
AR is supported in part by NSF grants PHY-0970173. 
The work of TMPT is supported in part by NSF grant PHY-0970171.


\begin{thebibliography}{99}

\bibitem{Ackermann:2011wa}
  M.~Ackermann {\it et al.}  [Fermi-LAT Collaboration],
  Phys.\ Rev.\ Lett.\  {\bf 107}, 241302 (2011)
  [arXiv:1108.3546 [astro-ph.HE]];

\bibitem{Abdo:2010nc}
  A.~A.~Abdo {\it et al.}  [ The Fermi-LAT Collaboration],
  Phys.\ Rev.\ Lett.\  {\bf 104}, 091302 (2010)
  [arXiv:1001.4836 [astro-ph.HE]].

\bibitem{Fermi:2012}
  M.~Ackermann {\it et al.}  [Fermi-LAT Collaboration],
  arXiv:1205.2739 [astro-ph.HE].


\bibitem{Profumo:2012tr}
  S.~Profumo and T.~Linden,
  arXiv:1204.6047 [astro-ph.HE].


\bibitem{Bringmann:2012vr}
  T.~Bringmann, X.~Huang, A.~Ibarra, S.~Vogl and C.~Weniger,
  arXiv:1203.1312 [hep-ph].

\bibitem{Weniger:2012tx}
  C.~Weniger,
  arXiv:1204.2797 [hep-ph].

\bibitem{Tempel:2012ey}
  E.~Tempel, A.~Hektor and M.~Raidal,
  arXiv:1205.1045 [hep-ph].
  
  
\bibitem{Goodman:2010qn}
  J.~Goodman, M.~Ibe, A.~Rajaraman, W.~Shepherd, T.~M.~P.~Tait and H.~-B.~Yu,
  Nucl.\ Phys.\ B {\bf 844}, 55 (2011)
  [arXiv:1009.0008 [hep-ph]].
  
\bibitem{Abazajian:2011tk}
  K.~N.~Abazajian, P.~Agrawal, Z.~Chacko and C.~Kilic,
  arXiv:1111.2835 [hep-ph].
  

  \bibitem{Bergstrom:1997fh}
  L.~Bergstrom and P.~Ullio,
  Nucl.\ Phys.\  B {\bf 504}, 27 (1997)
  [arXiv:hep-ph/9706232];
  Z.~Bern, P.~Gondolo and M.~Perelstein,
  Phys.\ Lett.\  B {\bf 411}, 86 (1997)
  [arXiv:hep-ph/9706538];
  P.~Ullio and L.~Bergstrom,
  Phys.\ Rev.\  D {\bf 57}, 1962 (1998)
  [arXiv:hep-ph/9707333];
 L.~Bergstrom, P.~Ullio and J.~H.~Buckley,
 Astropart.\ Phys.\  {\bf 9}, 137 (1998)
 [arXiv:astro-ph/9712318];
  F.~Boudjema, A.~Semenov and D.~Temes,
  Phys.\ Rev.\  D {\bf 72}, 055024 (2005)
  [arXiv:hep-ph/0507127].

\bibitem{Gustafsson:2007pc}
  M.~Gustafsson, E.~Lundstrom, L.~Bergstrom and J.~Edsjo,
  Phys.\ Rev.\ Lett.\  {\bf 99} (2007) 041301
  [arXiv:astro-ph/0703512];
  M.~Perelstein and A.~Spray,
  Phys.\ Rev.\  D {\bf 75} (2007) 083519
  [arXiv:hep-ph/0610357].

\bibitem{Bergstrom:2004nr}
  L.~Bergstrom, T.~Bringmann, M.~Eriksson and M.~Gustafsson,
  JCAP {\bf 0504}, 004 (2005)
  [hep-ph/0412001].

\bibitem{cline:2012}
  J.~M.~Cline,
  arXiv:1205.2688 [hep-ph];
  K.~-Y.~Choi and O.~Seto,
  arXiv:1205.3276 [hep-ph].

\bibitem{Bertone:2010fn}
  G.~Bertone, C.~B.~Jackson, G.~Shaughnessy, T.~M.~P.~Tait and A.~Vallinotto,
  JCAP {\bf 1203}, 020 (2012)
  [arXiv:1009.5107 [astro-ph.HE]].


\bibitem{Jackson:2009kg}
  C.~B.~Jackson, G.~Servant, G.~Shaughnessy, T.~M.~P.~Tait and M.~Taoso,
  JCAP {\bf 1004}, 004 (2010)
  [arXiv:0912.0004 [hep-ph]].
  

\bibitem{Bertone:2009cb}
  G.~Bertone, C.~B.~Jackson, G.~Shaughnessy, T.~M.~P.~Tait and A.~Vallinotto,
  Phys.\ Rev.\ D {\bf 80}, 023512 (2009)
  [arXiv:0904.1442 [astro-ph.HE]].

  

\bibitem{McDermott:2010pa}
  S.~D.~McDermott, H.~-B.~Yu and K.~M.~Zurek,
  Phys.\ Rev.\ D {\bf 83}, 063509 (2011)
  [arXiv:1011.2907 [hep-ph]].
  

\bibitem{Dudas:2009uq}
  E.~Dudas, Y.~Mambrini, S.~Pokorski and A.~Romagnoni,
  JHEP {\bf 0908}, 014 (2009)
  [arXiv:0904.1745 [hep-ph]].
  
\bibitem{Mambrini:2009ad}
  Y.~Mambrini,
  JCAP {\bf 0912}, 005 (2009)
  [arXiv:0907.2918 [hep-ph]].
  
\bibitem{Dudas:2012pb}
  E.~Dudas, Y.~Mambrini, S.~Pokorski and A.~Romagnoni,
  arXiv:1205.1520 [hep-ph].
  

\bibitem{Yang:1950rg}
  L.~D.~Landau, Dokl. Akad. Nawk., USSR {\bf 60}, 207 (1948);
  C.~N.~Yang,
  Phys.\ Rev.\  {\bf 77}, 242 (1950).
  
\bibitem{fermistat}
  W. A.~Rolke, A. M.~Lopez, and J.~Conrad, Nucl. Instrum. Meth. A551 (2005) 493–503.


\end{thebibliography}
\end{document}